\documentclass[reprint,aps,prl,superscriptaddress]{revtex4-2}

\usepackage[utf8]{inputenc}
\usepackage[T1]{fontenc}
\usepackage{amsmath,amssymb}
\usepackage{graphicx}
\usepackage{bm}
\usepackage{float}
\usepackage{xcolor}
\usepackage{ulem}
\usepackage{makecell}

\begin{document}

\title{
Acoustic Plasmon Resonance: Breaking the Anderson Stiffness Paradigm in Quasi-Two-Dimensional Superconducting Films}

\author{V.~M.~Kovalev}
\affiliation{Rzhanov Institute of Semiconductor Physics, Siberian Branch\\ of Russian Academy of Science, Novosibirsk 630090, Russia}
\affiliation{Novosibirsk State Technical University, Novosibirsk 630073, Russia}

\author{A.~V.~Chaplik}
\affiliation{Rzhanov Institute of Semiconductor Physics, Siberian Branch\\ of Russian Academy of Science, Novosibirsk 630090, Russia}
\affiliation{Novosibirsk State University, Novosibirsk 630090, Russia}

\date{\today}

\begin{abstract}
Recent experiments on superconducting films have revealed an acoustic plasmon mode that depends critically on the superconducting transition, directly challenging the long-standing Anderson-Higgs paradigm regarding the stiffness of the plasma spectrum in superconductors. In this Letter, we provide a microscopic theoretical framework that explains this behavior and establishes the physical conditions under which classical Anderson-Higgs constraints are bypassed. We demonstrate that in films of finite thickness, the transverse redistribution of normal and superfluid charge densities enables a unique coupling mechanism to electromagnetic radiation -- a feature fundamentally absent in the conventional Carlson-Goldman scenario. Our theory predicts an acoustic mode whose dispersion, temperature scaling, and dependence on film thickness are in remarkable agreement with recent experimental observations. By delineating the regime of this acoustic response, we reconcile the observed electromagnetic activity of collective excitations with the fundamental principles of superconductivity.
\end{abstract}

\maketitle

\textit{Introduction.}---The dynamics of collective excitations in low-dimensional superconductors is central to understanding quantum coherence and electromagnetic (EM) response in condensed matter systems. Since the seminal work by Anderson \cite{anderson}, it has been well established that long-range Coulomb interactions typically shift plasma oscillations to high frequencies. In three-dimensional (3D) systems, this ``stiffness'' of the plasma spectrum ensures that the plasmon frequency remains much larger than the superconducting gap $\omega_p \gg \Delta$, rendering the collective response largely insensitive to the superconducting transition. This robustness of the plasma spectrum extends to lower-dimensional systems, including two-dimensional, layered and one-dimensional electron gases, where the plasma frequency can be smaller than the energy gap, $\omega_p \ll \Delta$, but, nevertheless, the influence of the superconducting order parameter on high-frequency plasma charge dynamics is negligible \cite{ohashi}.

The only widely recognized exception near the transition temperature $T_c$ has been the Carlson-Goldman (CG) mode \cite{CG1, CG2}. This collective excitation arises from out-of-phase oscillations between the superconducting condensate and the normal fluid \cite{Schmid1975, Artemenko1978}. However, the physical nature of the CG mode imposes severe observational constraints. First, as a purely internal redistribution of density between the two components, it is inherently quasi-neutral. This neutrality makes it effectively ``invisible'' to direct interaction with external EM radiation; consequently, the CG mode has been detected only in specialized tunneling experiments. Second, standard models predict a low phase velocity ($v_{CG} \ll v_F$), a specific temperature scaling $\propto (T_c - T)^{1/4}$, and, crucially, an independence of the film thickness $d$, allowing the mode to persist even in the limit $d \to 0$.

Nevertheless, recent optical and microwave spectroscopy studies have uncovered an acoustic plasma mode that radically deviates from this classical paradigm \cite{Andreeva}. Unlike standard plasmons, this mode exhibits a critical dependence on the proximity to $T_c$, with its frequency residing deep within the superconducting gap ($\omega < \Delta$), and possessing low damping. In stark contrast to the CG mode, the observed excitation possesses a significant dipole moment, couples actively to the EM field, and is characterized by a high velocity ($s > v_F$). Furthermore, experiments reveal a fundamental role of geometry: the mode velocity follows a square-root scaling $\sqrt{T_c - T}$ and directly relates to the film thickness $d$, vanishing as the thickness approaches the strictly 2D limit. This phenomenon represents a new class of collective response that simultaneously exhibits plasmonic features and high sensitivity to the superconducting order, directly challenging the traditional stiffness of the plasma spectrum in superconducting structures.

In this Letter, we present a microscopic theory that resolves this fundamental paradox. We demonstrate that in superconducting films of finite thickness, a specific transverse distribution of normal and superconducting charge densities emerges. The spatial mismatch between these components across the film thickness $d$ leads to a breakdown of local neutrality and the emergence of an acoustic plasmon mode sensitive to the external EM radiation. Our analytical expressions for the frequency and damping quantitatively reproduce the experimental data over a wide temperature range, $0<T<T_c$. Our findings reveal that the transverse charge structure acts as a key mechanism allowing plasma modes to ``feel'' the superconducting phase, opening new avenues for the resonant manipulation of quantum condensates in low-dimensional systems.

\textit{Model.}---We consider a superconducting film of thickness $d$, where transverse confinement significantly modifies the electronic states. In the regime where the characteristic frequency of the plasma branch is small compared to the superconducting gap ($\omega<\Delta$), the collective response is governed by the subgap dynamics of both the superfluid and normal components. Impurity scattering is characterized by the normal-state momentum relaxation time $\tau$. Depending on the hierarchy of characteristic scales governing electron motion and correlations, we distinguish two transport regimes: (i) the clean limit ($l \gg \xi_0$) and (ii) the dirty limit ($l \ll \xi_0$), where $l=v_F\tau$ denotes the mean free path and $\xi_0$ is the superconducting coherence length. In particular, the experimental conditions reported in Ref.~\cite{Andreeva} correspond to the diffusive (dirty) limit at low frequencies, $\omega\tau \ll 1$.

To describe regime (i), we employ the kinetic equation approach for superconductors \cite{AronovGurevich}, while the dirty limit (ii) is treated within the framework of the Mattis-Bardeen theory \cite{MattisBardeen}. In the following, we provide a detailed analysis of each transport regime.

The kinetic equation approach provides a self-consistent treatment of the non-equilibrium quasiparticle distribution and its coupling to the order parameter phase. The dynamics of the quasiparticle distribution function $n_{\bf p}({\bf r},t)$ and the condensate momentum ${\bf p}_s$ are governed by:
\begin{gather}\label{AG}
\partial_t {\bf p}_s - \partial_{\bf r} \Phi= e{\bf E},
\\\nonumber
\partial_t n({\bf p},{\bf r},t) + \partial_{\bf p}\tilde{\epsilon}_{\bf p} \cdot \partial_{\bf r}n({\bf p},{\bf r},t)
\\\nonumber 
- \partial_{\bf r}\tilde{\epsilon}_{\bf p} \cdot \partial_{\bf p}n({\bf p},{\bf r},t) = Q\{n({\bf p},{\bf r},t)\},
\end{gather}
where $Q\{n({\bf p},{\bf r},t)\}$ is the collision operator for quasiparticles scattering off impurities. The local quasiparticle energy is $\tilde{\epsilon}_{\bf p} = \epsilon_{\bf p} + {\bf v}\cdot{\bf p}_s$, with ${\bf v}={\bf p}/m$, $\epsilon_{\bf p} = \sqrt{\tilde{\zeta}_{\bf p}^2 + \Delta^2}$, $\tilde{\zeta}_{\bf p} = \zeta_{\bf p} + \Phi + p_s^2/2m$, and $\zeta_{\bf p} = p^2/2m - \mu$, with $\mu$ being Fermi energy. The gauge-invariant potentials are defined as ${\bf p}_s = \frac{1}{2}\nabla\chi - \frac{e}{c}{\bf A}$ and $\Phi = e\phi + \frac{1}{2}\partial_t\chi$, where $\phi$ and ${\bf A}$ are the scalar and vector potentials, respectively, and $\chi$ is the phase of the order parameter.
For point-like impurities with a potential $U_{\bf p,p'}=U_0$, the collision operator
\begin{gather*}
Q\{n_{\bf p}({\bf r},t)\}=-2\pi N_i\sum_{\bf p}|U_{\bf p,p'}|^2(u_{\bf p}u_{\bf p'}-v_{\bf p}v_{\bf p'})^2\times\\\nonumber
\times\delta(\tilde{\epsilon}_{\bf p}-\tilde{\epsilon}_{\bf p'})
[n({\bf p})-n({\bf p}')],\,\,\,
\begin{pmatrix}
  u^2_{\bf p} \\
  v^2_{\bf p}
\end{pmatrix}=\frac{1}{2}\left(1\pm\frac{\tilde{\zeta}_{\bf p}}{\epsilon_{\bf p}}\right),
\end{gather*}
%
%
%
%
simplifies to the relaxation-time approximation $Q\{n_{\bf p}\} = - \delta n({\bf p}) / \tau_p$, where the energy-dependent relaxation rate is $\tau_p^{-1} = \tau^{-1} |\zeta_{\bf p}| / \epsilon_{\bf p}$, and $\tau^{-1}=mN_i|U_0|^2$ is the momentum relaxation time in the normal state, with $N_i$ being the impurity density. It should be noted that when the scattering probability is energy-dependent (scattering off Coulomb centers, etc.), the relaxation time may be evaluated at the Fermi level, such that $\tau^{-1}=\tau^{-1}(\mu)$.

A key feature of our model is the explicit treatment of the transverse charge distribution for both Cooper pairs and normal excitations. As the system deviates from the idealized 2D limit, plasma oscillations involve dynamic fluctuations of the bulk charge density, inducing long-range Coulomb forces. In the quasi-static regime, the induced scalar potential $\phi$ satisfies the Poisson equation:
\begin{gather}\label{poisson}
\left(\frac{d^2}{dz^2}-k^2\right)\phi_{k\omega}(z) = 
\\\nonumber
=-4\pi e \Bigl[ A_o(z)\delta N_C
+A_s(z)\delta N^s_{k\omega} + A_n(z)\delta N^n_{k\omega} \Bigr].
\end{gather}
Here, $A_{\alpha}(z)$ (where $\alpha = s, n, o$) are charge distribution functions normalized across the film thickness, satisfying $\int A_\alpha(z) dz = 1$ normalization condition. The terms $\delta N^{s,n}_{k\omega}$ denote the non-equilibrium perturbations of the superfluid and normal densities, respectively, while $\delta N_0$ represents
the contribution of electrons residing deeply under Fermi level. The coupling of the former two quantities with the external potential is determined by the kinetic equation or Mattis-Bardeen approach (see below), whereas the latter, in the quasi-static limit, takes the form $\delta N_0=(\partial N/\partial \mu) e\phi_{k\omega}$, with $N$ being the total electron density.

Equation \eqref{poisson} is supplemented by the continuity equations for the total charge and current densities. We assume that the continuity equation holds for each component individually; this approach ensures global charge conservation and is justified by the suppression of pair-breaking processes in the adiabatic limit, $\omega\ll\Delta$.

Linearizing the kinetic equation \eqref{AG} for small perturbations $\propto \exp(-i\omega t + i{\bf kr})$, we obtain the first-order correction to the distribution function:
\begin{equation}\label{firstorder}
\delta n^{(1)}_{k\omega}({\bf p}) = \omega \frac{{\bf v} \cdot {\bf p}^s_{k\omega} + \Phi_{k\omega} \zeta/\epsilon}{\omega - {\bf kv}\zeta/\epsilon + i/\tau_{p}} (-n_0'),
\end{equation}
where $n_0=[\exp(\epsilon_{\bf p}/T)+1]^{-1}$ is the equilibrium quasiparticles distribution function. 
This allows us to calculate the current densities ${\bf j}^s_{k\omega} = e N_s {\bf p}^s_{k\omega}/m$ and ${\bf j}^n_{k\omega}=e\sum_{\bf p}{\bf v}\delta n^{(1)}_{k\omega}({\bf p})$, where
\begin{gather}\label{currents}
{\bf j}^n_{k\omega}=e\sum_{\bf p}
\frac{\omega{\bf v}(-n_0')}
{\omega-{\bf kv}\zeta/\epsilon+i/\tau_{p}}\left({\bf v}\cdot{\bf p}^s_{k\omega}+\Phi_{k\omega}\frac{\zeta}{\epsilon}\right)
\end{gather}
By self-consistently expressing the potentials ${\bf p}^s_{k\omega}$ and $\Phi_{k\omega}$ by means of the continuity equations via the electric field ${\bf E}$, 
\begin{gather}\label{currents2}
\Phi_{k\omega}=\frac{-ieV^2_{k\omega}({\bf kE})}{\omega^2-k^2V^2_{k\omega}},\,\,{\bf p}^s_{k\omega}=\frac{ie\omega{\bf E}}{\omega^2-k^2V^2_{k\omega}}.
\end{gather}
we arrive at the general dispersion relation for collective modes:
\begin{gather}\label{dispersion}
1 - U_k (\Pi^s_{k\omega} J_{ss} + \Pi^n_{k\omega} J_{nn} + \Pi^0 J_{oo})+\\\nonumber
+ U_k^2 (\Pi^s_{k\omega} \Pi^n_{k\omega} M_{sn} + \Pi^s_{k\omega} \Pi^0 M_{so} + \Pi^n_{k\omega} \Pi^0 M_{no}) 
\\\nonumber
- U_k^3 \Pi^s_{k\omega} \Pi^n_{k\omega} \Pi^0 \det |J| = 0.
\end{gather}
Here, $U_k = 2\pi e^2/k$ denotes the 2D Coulomb potential (the static dielectric permittivity of the environment can be accounted for via the substitution $e^2 \rightarrow e^2/\varepsilon$, where $\varepsilon = (\varepsilon_0 + 1)/2$ for a film sandwiched between a substrate with permittivity $\varepsilon_0$ and a vacuum, as in \cite{Andreeva}), $M_{ij} = J_{ii} J_{jj} - J_{ij}^2$,
\begin{gather}\nonumber
\det |J| = J_{ss}(J_{nn}J_{oo}-J_{no}^2) - J_{sn}(J_{sn}J_{oo} - J_{so}J_{no})+\\
+ J_{so}(J_{sn}J_{no} - J_{nn}J_{so}).
\end{gather}
The form factors are defined as
\begin{gather}\label{formfactors}
J_{\alpha\beta}=\int dz\int dz'A_\alpha(z)e^{-k|z-z'|}A_\beta(z'),
\end{gather}
that account for the transverse charge geometry. The polarization operators $\Pi^{s,n}_{k\omega}$ yield
\begin{gather}\label{polarization_operators}
\Pi^s_{k\omega}=\frac{k^2N_s/m}{\omega^2-k^2V^2_{k\omega}},\,\,\,\,\Pi^0=\partial N/\partial\mu,\\\nonumber
\Pi^n_{k\omega}=\sum_{\bf p}\frac{({\bf kv})}{\omega^2-k^2V^2_{k\omega}}
\frac{\omega({\bf kv})-k^2V^2_{k\omega}\zeta/\epsilon}{\omega-({\bf kv}\zeta/\epsilon)+i/\tau_p}(-n_0').
\end{gather}
Here, the equilibrium superconducting density reads $N_s=N(1-f_T)$, where $f_T=2\int_0^\infty(-n_0')d\zeta$, and $N=m\mu/(2\pi)$ is the total electron density. 
Generally, Eq.~(\ref{dispersion}) yields two dispersion branches of the superconducting film and their respective damping rates; however, the experiment in Ref.~\cite{Andreeva} addresses exclusively the low-frequency regime ($\omega\tau\ll1$).

In the clean limit ($l \gg \xi_0$), where the evolution of both normal and superconducting components is described by the kinetic equation, the low-frequency dynamics ($\omega\tau \ll 1$) are characterized by an overdamped response of the normal component, while the superconducting component remains oscillatory. In this limit, the normal and superconducting component polarization operators are approximated as
\begin{gather}\label{moderate1}
\Pi_{k\omega}^{s} = \frac{k^2 N_s}{m\omega^2},\\\nonumber
\Pi^n_{k\omega}\approx\frac{-i}{\omega}\sum_{\bf p}({\bf kv})^2\tau_p(-n_0')=
\frac{k^2N_n}{m\omega^2}(-i\omega\tau),
\end{gather}
where we introduced the normal density as $N_n=2N\int_0^{\infty} d\zeta (\epsilon/|\zeta|)(-n_0')$.

In the dirty limit, $l \ll \xi_0$, the kinetic equation approach is inapplicable. This regime requires a description based on the conductivities of the normal and superconducting fractions under strong scattering. In this regime, we apply the Mattis-Bardeen approach \cite{MattisBardeen}. Since the polarization operators are directly related to the complex conductivities via the continuity equation, they can be expressed in terms of the Mattis-Bardeen conductivities \cite{MattisBardeen} $\sigma_{s,n}(\omega)$:
\begin{gather}\label{Paperpolarisationdurty}
\Pi^{s}_{k\omega} = \frac{k^2 \sigma_s(\omega)}{e^2 \omega} = \frac{k^2 \tilde{N}_s}{m\omega^2}, \\\nonumber \Pi^{n}_{k\omega} = -i\frac{k^2 \sigma_n(\omega)}{e^2 \omega} = \frac{k^2 \tilde{N}_n}{m\omega^2}(-i\omega\tau),
\end{gather}
and the same expression for $\Pi_0$ as above. The densities in dirty regime are introduced as ($\omega<\Delta$):
\begin{gather}\label{PaperDensities}
\tilde{N}_s(\omega) = N\tau \int_{\Delta-\omega}^{\Delta} d\epsilon \frac{[1-2n_0(\epsilon+\omega)][\epsilon^2+\Delta^2+\epsilon\omega]}{\sqrt{\Delta^2-\epsilon^2}\sqrt{(\epsilon+\omega)^2-\Delta^2}},
\\\nonumber
\tilde{N}_n(\omega) =\frac{2N}{\omega} \int_{\Delta}^{\infty} d\epsilon \frac{[n_0(\epsilon)-n_0(\epsilon+\omega)][\epsilon^2+\Delta^2+\epsilon\omega]}{\sqrt{\epsilon^2-\Delta^2}\sqrt{(\epsilon+\omega)^2-\Delta^2}}.
\end{gather}
Comparing relations \eqref{Paperpolarisationdurty} with \eqref{moderate1}, we observe that they are identical, differing only in the definitions of the densities $N_s$ and $N_n$.

\textit{Collective mode.}---Substituting \eqref{moderate1} (or \eqref{Paperpolarisationdurty}) into the general dispersion equation \eqref{dispersion}, we obtain:
\begin{gather}
1 + \frac{J_{oo}}{ka} - \frac{\omega_0^2 N_s}{\omega^2 N} \left( J_{ss} + \frac{M_{so}}{ka} \right) =
\\\nonumber
=-i\omega\tau \frac{\omega_0^2 N_n}{\omega^2 N} \left[ J_{nn} + \frac{M_{no}}{ka} + \frac{\omega_0^2 N_s}{\omega^2 N} \left( \frac{\det|J|}{ka} - M_{sn} \right) \right], 
\end{gather}
which, after algebraic simplification, takes the form:
\begin{gather}\label{PaperDispersion}
\omega^2 - \omega_0^2 \frac{N_s}{N} \frac{M_{so} + ka J_{ss}}{J_{oo} + ka} =-i\omega\tau \omega_0^2 \frac{N_n/N}{J_{oo} + ka} \times
\\\nonumber
\times\left[ ka J_{nn} + M_{no} + \frac{\omega_0^2 N_s}{\omega^2 N} (\det|J| - ka M_{sn}) \right]. 
\end{gather}
Here, $a^{-1} = \pi e^2\partial N/\partial \mu$ represents an effective screening length (Bohr radius), and $\omega_0^2 = 2\pi e^2 k N / m$ is the plasmon frequency for a 2D electron gas. Eq. \eqref{PaperDispersion} is formally a full a full cubic equation  relative to $\omega$. We concentrate in what follows on one of its solutions describing the acoustic plasmon mode which has been observed in experiments \cite{Andreeva}. Assuming a weakness of imaginary part due to smallness of $\omega\tau \ll 1$, we solve \eqref{PaperDispersion} using successive approximations. The resulting solution is $\omega = \omega_k - i\Gamma$, with mode frequency $\omega_k$ and damping rate $\Gamma$ are:
\begin{gather}\label{Papersolutions}
\omega_k = \omega_0 \sqrt{\frac{N_s}{N} \frac{M_{so} + ka J_{ss}}{J_{oo} + ka}}, 
\,\,\,\,
\Gamma = \frac{\omega_0^2 \tau_i}{2} \frac{N_n}{N} S(k), 
\end{gather}
where the structure factor $S(k)$ is defined as:
\begin{equation}
S(k) = \frac{M_{no} + ka J_{nn}}{J_{oo} + ka} + \frac{\det|J| - ka M_{sn}}{M_{so} + ka J_{ss}}.
\end{equation}
%
%
%
%
%
%
The dependence of the mode frequency and damping on the wavevector $k$ is determined by the geometric form factors $J_{\alpha\beta}(k)$ and the effective length $a$. 
The scaling of $J_{\alpha\beta}(k)$, $M_{\alpha\beta}(k)$, and $\det|J|$ with $k$ depends on the mutual relations among the functions $A_\alpha(z)$. 
Using experimental data~\cite{Andreeva}, we estimate $a \sim 0.3\,\text{nm} \ll d = 8\,\text{nm}$, which yields the long-wavelength behavior of the structure factors:
\begin{gather}\label{factors} 
\frac{M_{so} + ka J_{ss}}{J_{oo} + ka}\approx D_0kd,\quad S(k) \approx S_0, 
\end{gather}
where $D_0$ and $S_0$ are constant dimensionless parameters. 
Consequently, the frequency exhibits an acoustic-type dispersion law,
\begin{equation}\label{DispersionDamping}
\omega_k = sk = \omega_0 \sqrt{D_0\frac{N_s}{N} kd},\quad \Gamma = \frac{\omega_0^2 \tau}{2} \frac{N_n}{N} S_0. 
\end{equation}

This linear dispersion describes an acoustic collective mode with weak damping provided $\omega_0\tau\ll 1$. In dirty case the normal $\tilde{N}_n$ and superconducting $\tilde{N}_s$ densities should be taken at low frequency,  $\omega\ll\Delta$, according to experiment \cite{Andreeva}. The phase velocity scales as $s \propto \sqrt{d}$. If the bulk $3D$ total concentration of electrons of the finite width film $N_V$ is fixed as a material parameter, then the areal concentration $N$ equals $N_Vd$ and our theory predicts $\omega_k\propto d$. This statement has been experimentally verified for three different values of thickness $d=5\,nm,\,8\,nm$ and $d=12\,nm$. A good qualitative agreement with the theory has been established \cite{Kukushkin}.

%
%
%
%
%
%
%
%
%

We now examine the qualitative behavior of the emergent acoustic mode. Physically, this mode characterizes condensate oscillations, while the normal component remains overdamped, as evidenced by the polarization operators in Eqs.~\eqref{moderate1} and \eqref{Paperpolarisationdurty}, thereby causing damping of the acoustic branch. The temperature dependence of both the mode velocity and its damping is strictly governed by the thermal evolution of the superconducting and normal fluid densities. The temperature behavior of the superconducting density in the limits of low and high temperatures are presented in Table \ref{tab:ns_simple}. 
\begin{table}[b]
\centering
\caption{Relative superconducting density, $N_s(T)/N$. The data in the table are the limiting cases of the general expression $N_s(T)/N=1-f_T$ for the clean case and $\tilde{N}_s(T)/N=\pi\Delta\tau_i\tanh(\Delta/2T)$ for a dirty one, coming from \eqref{PaperDensities} at $\omega\ll\Delta$.}
\label{tab:ns_simple}
\renewcommand{\arraystretch}{2.5}
\begin{tabular}{|l|c|c|}
\hline
\textbf{Limit} & \textbf{Clean} ($l \gg \xi_0$) & \textbf{Dirty} ($l \ll \xi_0$) \\ \hline
\begin{tabular}[c]{@{}l@{}}
$T \ll\Delta$ \end{tabular} & $\displaystyle \frac{N_s}{N} \approx 1 - \sqrt{\frac{2\pi\Delta}{T}} e^{-\frac{\Delta}{T}}$ & $\displaystyle \frac{\tilde{N}_s}{N} \approx \pi \tau_i \Delta \left( 1 - \sqrt{\frac{2\pi\Delta}{T}} e^{-\frac{\Delta}{T}} \right)$ \\ \hline
\begin{tabular}[c]{@{}l@{}}
$T\gg\Delta$\end{tabular} & $\displaystyle \frac{N_s}{N} \approx \frac{7\zeta(3)}{4\pi^2} \frac{\Delta^2}{T^2}$ & $\displaystyle \frac{\tilde{N}_s}{N} \approx \frac{\pi T\tau_i}{2} \frac{\Delta^2}{T^2}$ \\ \hline
\end{tabular}
\end{table}
In both the clean and dirty limits, the acoustic mode velocity approaches a constant value at low temperatures. Conversely, in the vicinity of $T_c$, the mode velocity in clean regime behaves as:
\begin{equation} \label{clean_crit}
s=\sqrt{\frac{2\pi e^2Nd}{\varepsilon m}\left(\frac{7\zeta(3)}{4\pi^2} \frac{\Delta^2}{T^2}\right)} \propto \sqrt{T_c - T},
\end{equation}
where $\zeta(x)$ is a Riemann zeta function.  In the dirty limit the phase velocity reads:
\begin{equation} \label{dirty_crit}
s=\sqrt{\frac{2\pi e^2Nd}{\varepsilon m}\left(\frac{\pi \tau \Delta^2}{2T}\right)} \propto \sqrt{(\tau T_c)(T_c - T)}.
\end{equation}
Notably, the acoustic mode exhibits a universal $\sqrt{T_c - T}$ scaling law near the critical temperature $T_c$ within both regimes. 
However, in disordered samples, this mode is reduced by a factor of $\tau T_c \ll 1$. 
The acoustic branch remains well-defined even under conditions where the dynamics of the normal component are purely overdamped. 
This temperature dependence is in excellent agreement with the experimental data reported by Andreeva et al.~\cite{Andreeva}.
\begin{figure*}[t]
    \centering
    \includegraphics[width=0.49\textwidth]{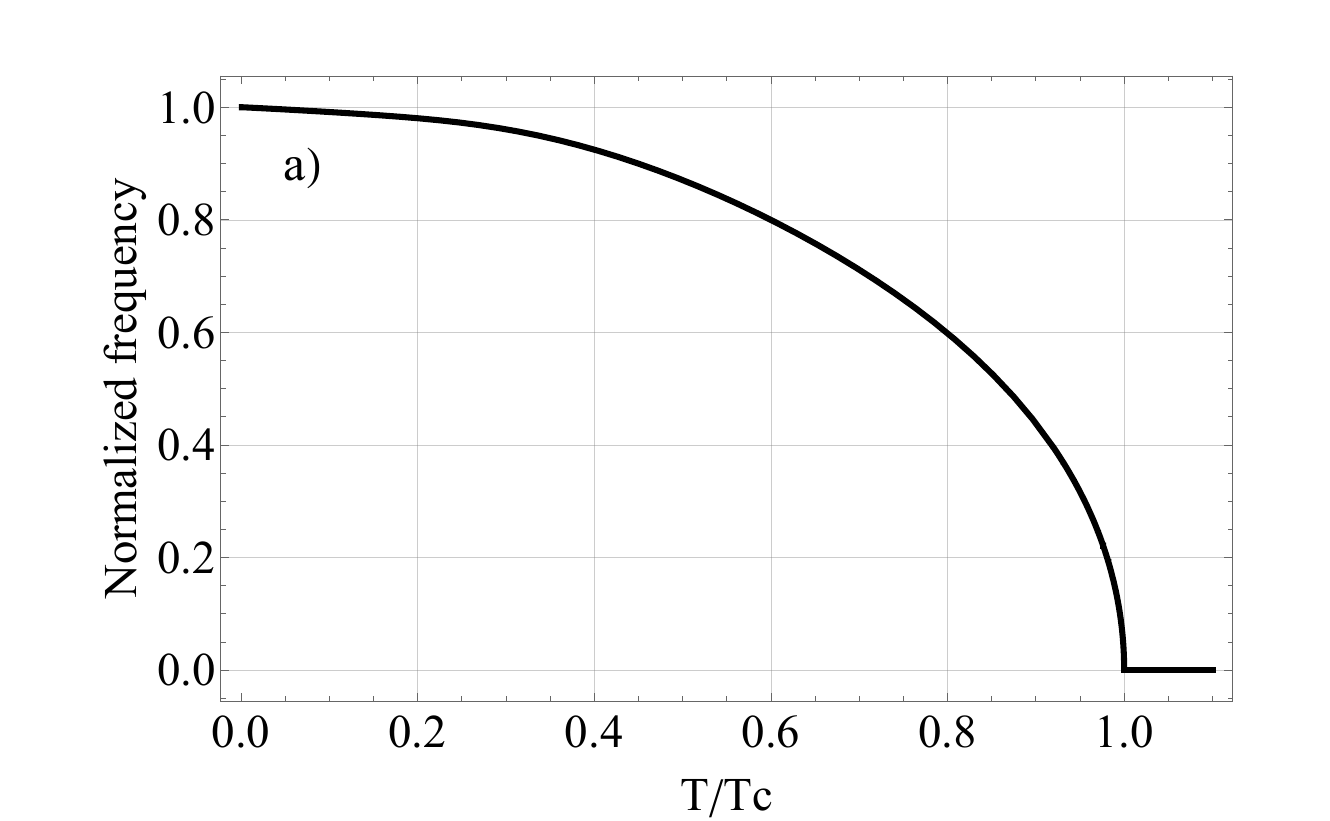}
    \hfill
    \includegraphics[width=0.49\textwidth]{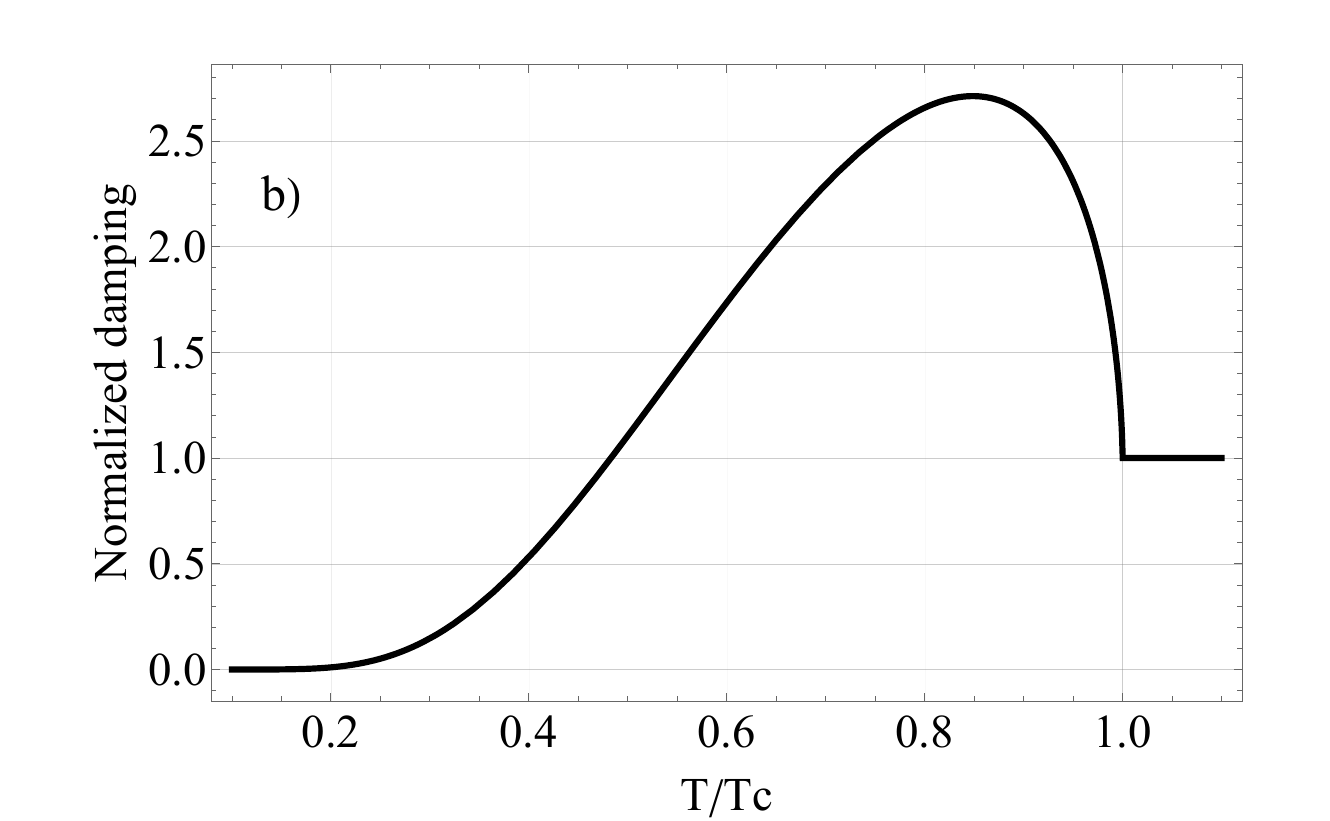}
    \caption{Acoustic mode properties as a function of temperature in the dirty regime. 
    (a) Temperature dependence of the normalized acoustic mode dispersion, $\omega_k(T)/\omega_k(0)$, versus the reduced temperature, $T/T_c$. 
    (b) Temperature dependence of the normalized acoustic mode damping, $\Gamma(T)/\Gamma(T_c)$, versus the reduced temperature, $T/T_c$. The phenomenological relation $\Delta(T)=\Delta(0)\tanh\left(1.74\sqrt{T_c/T-1}\right)$ for a dependence $\Delta(T)$ in all region $0<T<T_c$ was used. 
    }
    \label{fig:plots}
\end{figure*}

The damping of the acoustic mode is determined by the temperature dependence of the normal components, $N_n$ and $\tilde{N}_n$, in the clean and disordered regimes, respectively. In both regimes, these components become negligibly small in the low-temperature limit $T \rightarrow 0$ and approach the total density $N$ at the critical temperature $T = T_c$. 
The calculated temperature dependencies of the normalized acoustic frequency $\omega_k(T)/\omega_k(0)$ and normalized damping $\Gamma(T)/\Gamma(T_c)$ in dirty regime in the temperature range $0<T<T_c$ are presented in Fig.~\ref{fig:plots}. The numerical calculations were performed using the following parameter set borrowed from the experimental structure \cite{Andreeva}: quantum well width $d=8\,nm$, 2D total electron density $N=10^{17}\,cm^{-2}$, $m\approx m_0$, $\varepsilon_0=11.5$, relaxation time $\tau=10^{-15}\,s$, critical temperature $T_c=13.7\,K$ and wavevector $k\approx 80\,cm^{-1}$. These data give the estimations $\omega_k(T=0)\approx 113$ GHz, what corresponds to $s(T=0)\approx1.4\cdot 10^9\,cm/s$ for $D_0\sim70$. The damping $\Gamma(T_c)\approx 7.4$ GHz for $S_0\sim7.4\cdot10^{-3}$, and superconducting density in dirty limit $\tilde{N}_s(T=0)\approx 1.4\cdot10^{14}\,cm^{-2}$.

The numerical analysis of acoustic mode frequency and its damping as functions of temperature is 
presented in Fig.~\ref{fig:plots}. The frequency decreases with increasing temperature 
and vanishes above $T_c$. Conversely, the damping increases with temperature, reaching 
the normal-state value above $T_c$. The latter corresponds to the overdamped mode of a 
two-dimensional normal electron gas, $\omega_k = -i \omega_0^2 \tau$, normalized by the 
geometric factor $S_0$, as follows from Eq.~\eqref{DispersionDamping}, where $N_n = N$ above $T_c$. 
Before reaching the normal-state value at $T = T_c$, the temperature dependence of the 
damping exhibits a local maximum. This behavior is directly related to the Hebel-Slichter 
coherence peak observed in the nuclear magnetic resonance phenomenon in superconductors~\cite{HebelSlichter}. 
In general, the temperature dependencies presented in Fig.~\ref{fig:plots} are in reasonable 
agreement with the experimental observations reported in Ref.~\cite{Andreeva}.

The microscopic origin of the spatial variations of normal and superconducting densities across the film thickness remains an open question that extends beyond the scope of the current theoretical framework. Experimentally, such behavior may arise in strongly disordered samples where the superconducting coherence length becomes comparable to or smaller than the film thickness. 
A comprehensive understanding of these mechanisms requires further investigation and remains a subject for future study. Here we only note that in the plasma waves problem for a thin film with normal electron gas the spatial distribution of electrons features arise automatically and the functions $A_\alpha(z)$ are nothing but  the squared moduli of the transversal electron wave functions \cite{Vitlina}.

\textit{Conclusions.}---In summary, we have developed a microscopic kinetic theory of collective excitations in superconducting films of finite thickness, accounting for non-trivial transverse charge distributions and subgap quasiparticle dynamics. Our work identifies several key advancements and resolves discrepancies in the understanding of recent experiments \cite{Andreeva} observed acoustic plasmon response in quasi-two-dimensional superconducting films.

First, we have demonstrated that the traditional "stiffness" of the plasma spectrum, often invoked to argue for the insensitivity of plasmons to the superconducting transition, is fundamentally bypassed in films with finite thickness. By explicitly treating the spatial mismatch between the transverse profiles of the superfluid and normal components, we show that local charge neutrality is broken. This mechanism gives rise to the acoustic mode, enabling its strong coupling to external electromagnetic radiation; a feature traditionally thought to be absent for Carlson-Goldman acoustic-like modes in superconductors.

Second, our theory resolves the conflict between the classical Carlson-Goldman scenario and recent experimental observations \cite{Andreeva}. Unlike the CG mode, which is characterized by a low velocity ($v_{CG} \ll v_F$) and a $(T_c - T)^{1/4}$ scaling, the acoustic plasmon identified here exhibits a high phase velocity ($s > v_F$) and a robust $\sqrt{T_c - T}$ temperature dependence near $T_c$. Furthermore, we have established that this mode is intrinsically tied to the film geometry, vanishing in the idealized 2D limit ($d \to 0$), which provides a clear explanation for its emergence only in specific experimental architectures.


These results reconcile the observed electromagnetic activity of subgap collective excitations with the fundamental principles of superconductivity. Beyond explanation recent measurements, our findings open new avenues for the resonant manipulation of quantum condensates and provide a theoretical foundation for utilizing acoustic plasmons in the design of superconducting optoelectronic and sensing devices.

\textit{Acknowledgments.}---This work was supported by the Ministry of Science and Higher Education of the Russian Federation (Project FWGW-2025-0009), and the Foundation for the Advancement of Theoretical Physics and Mathematics ``BASIS''.

\end{document}